\documentclass{article}

\usepackage[american]{babel}

\title{Don't choose theories: Normative inductive reasoning and the status of physical theories}
\author{Andr\'e C. R. Martins \\
	NISC -- EACH -- Universidade de S\~ao Paulo,\\
	Rua Arlindo B\'etio, 1000, 03828--000,  S\~ao Paulo, Brazil}

\begin{document}


	\maketitle
	

	\begin{abstract}
		
		Evaluating theories in physics used to be easy. Our theories provided very distinct predictions. Experimental accuracy was so small that worrying about epistemological problems was not necessary. That is no longer the case. The underdeterminacy problem between string theory and the standard model for current possible experimental energies is one example. We need modern inductive methods for this problem, Bayesian methods or the equivalent Solomonoff induction. To illustrate the proper way to work with induction problems I will use the concepts of Solomoff induction to study the status of string theory. Previous attempts have focused on the Bayesian solution. And they run into the question of why string theory is widely accepted with no data backing it.  Logically unsupported additions to the Bayesian method were proposed.  I will show here that, by studying the problem from the point of view of the Solomonoff induction those additions can be understood much better. They are not ways to update probabilities. Instead, they are considerations about the priors as well as heuristics to attempt to deal with our finite resources. For the general problem, Solomonoff induction also makes it clear that there is no demarcation problem. Every possible idea can be part of a proper scientific theory. It is just the case that data makes some ideas extremely improbable. Theories where that does not happen must not be discarded. Rejecting ideas is just wrong.


	\end{abstract}

	%
	%
	%

	\section{Introduction}
	
	Describing physics as a successful history among sciences is too much of a common place. But it also carries some truth to it. Even more than fifty years ago, Wigner \cite{wigner60a} was already perplexed by the amazing agreement between theoretical predictions and experiments. To the point he considered our mathematical success as unreasonable. Perhaps even more surprising than the number of digits in the agreement between experiment and theory, we have had cases where we have discovered unexpected objects and entities. The existence of Neptune was predicted because the calculated orbit of Uranus did not match its observed behavior. Dirac predicted anti-particles as a second solution to his equations before they were observed. Physicists love those stories.  We all know them well, as great examples of the power of our calculations. Cautionary tales, such as the belief that light needed an aether to propagate, are often told as steps to a better theory and the lesson in caution, lost.
	
	And it is not in the precision of our theories that physicists can claim success. In most cases, our theories have provided us with predictions that, given the accuracy of our experiments, were also very discriminating. Newtonian Mechanics and General relativity provide almost identical predictions for many systems. But there are plenty of cases where the expected behavior is very different. And we can make experiments that show General Relativity is a better theory. This can be said with certainty or, more exactly, we can be so close to certainty that the difference is pure technicality. Chances smaller than 1 in $10^{100}$ can be called zero, without much abuse of language, after all. We have had success both in obtaining very precise theories, as well as very discriminating ones. Under these circumstances, epistemological considerations were not necessary. Everything works fine if you know epistemology well. But things also work fine if you know nothing about epistemological problems. We have been living in a epistemically safe world.
	
	But that is not always the case. As we look at the status of our current theories and how the community works with them, it is not hard to notice that there is something odd going on. We have never-ending discussions about the status of string theory. Those discussion even include the question of whether it is an acceptable theory \cite{ellissilk14a}. We currently can not make observations that would provide evidence that could allow us to choose between string theory and the standard model.  Based on that, some physicists claim string theory is not scientific. And science, we were taught, must be testable. And it must provide one single answer. Or must it not?

	Inductive reasoning, such as Bayesian methods, can help us answer that question \cite{jaynes03}. Indeed, recent efforts to bring some modern, non dated epistemology to the question \cite{dawid13a} have point out that the underdetermination problem is real and something we might have to learn to accept. Underdetermination happens when the predictions of two different models are identical. As a consequence, the probability ratio between the two models can not be altered by observing any data. And the initial, often subjective prior guesses remain. Under these circumstances, attempts to determine how to choose between underdetermined theories have caused deep disagreements between supporters and critics of String Theory \cite{castelvecchi15a}. In particular, the idea that a theory could be ``confirmed'' purely by theoretical arguments \cite{dawid06a} is a claim met with strong resistance. There is good reason for that. Bayesian methods can be obtained from logical considerations. And they have way to change probabilities from theoretical considerations. 
	
	In this paper, I will show why inductive methods are a necessity and which conclusions we can draw from them. I will explain why human cognition makes normative methods to choose theories a real necessity \cite{martins16b}. There are too many traps in our brains, we just can't trust our feelings to pick the best ideas. I will briefly explain that crisis in other areas, including the use of statistical tools, can also be traced to harmful interactions between our natural reasoning and our current tools \cite{martins16b}. We must use inductive methods. And we must understand them well and reason as close as possible to their demands. 
	
	Bayesian methods have some important consequences. Theories can no longer be discarded in the majority of cases. They can and must be ranked probabilistically. To further understand the consequences of inductive methods, I will see what else they can tell us about the status of string theory. I will show that both sides of the quarrel have captured important aspects of the problem. Instead  of using Bayesian induction to the ranking of theories, I will use an equivalent framework known as Solomonoff induction \cite{solomonoff64a}. This will provide us a new way to look at the theoretical considerations issue.  While we must keep all underdetermined theories as possibilities, we will see that the theoretical methods of ``confirmation'' do not provide ways to update probabilities. But they can carry important information to determine a priori assessments as well as carry some heuristical value for limited beings.

	\section{Our cognitive need for normative epistemology}
	
	When physicists try to predict the path of a newly found asteroid, they do not just look at the data and provide an educated guess. We all know better than to trust our brains when our equations do a much better job at it. And there are good reasons for that. Our brains did not evolve to track nor predict gravitational trajectories. And they also did not evolve to choose between competing theories or even ideas. Quite the opposite. Recent evidence shows that our reasoning skills might not have evolved to help us look for truth. Instead, they might exist to help us win arguments, regardless of truth \cite{merciersperber11a}. When thinking about identity-defining ideas, we use our analytical skills to defend our positions  \cite{kahan13a}.  And the better our skills are, the better we are at reasoning to find ways to defend ideas. Even when they are wrong. 
	
	We also have a tendency to be overconfident about our reasoning abilities \cite{oskamp65a}. But our confidence also plays strange tricks on us. As we get more data about an issue, we reach a point where our accuracy  in making predictions using that data stops improving.  At the same time, we keep getting more and more confident \cite{tsaietal08a}, despite no real improvement. Unfortunately, overconfidence is not limited to the issues we are not specialists. Quite the opposite. It is often the case that experts tend to consider themselves far better than they actually are \cite{christensenszalanskbusyhead81a}. We can be trained to know how good our estimates are, but that involves a lot of feedback about when we get things right. It usually also requires better models to support our estimates. Such an improvement in self calibration has happened, for example, in meteorological prediction \cite{murphywinkler84a}.

	In general,  actuarial judgments, where simple statistical models replace the human decision process, often provide a significant improvement to rate of successful evaluations \cite{dawesetal89a}. Our illusions about our own skills extend even to the point where we consider that random events, like the throw of a die, are more likely to benefit us if we make the throws ourselves \cite{proninetal06a}. We believe we have far more control over the world and our own reasoning than we actually do.
	
	The examples I just presented form a very small sample of a huge literature in how our natural reasoning fails. Our reasoning fails and, still, we trust our brains and our own evaluations when we should not. This means that, when we have a belief, we tend to be too confident about it. And, since argumentation might be more about convincing than getting the best answer, we are expected to defend our beliefs even when evidence is against us. Add to this the fact that logic and mathematics can prove nothing without initial axioms. No belief about the real world can be proven true. Deduction requires previous knowledge and inductive methods, while immensely helpful, do not provide certainty. Combining those facts, we can see that, when we believe, we harm our ability to look for truth (or best answers). We should actually hold no beliefs and, wherever possible, replace our own analysis for mathematical models \cite{martins16a}. That is exactly what we need to undestand the arguments in the String Theory debate better.

	\section{Bayes and Solomonoff: Moving beyond Popper}
	
	An outdated but still influential attempt on using logic to the problem of theory choice was the Popperian notion of falseability \cite{popper59}. Popper wanted to solve the demarcation problem, to find a way to divide theories in scientific and non-scientific ones. He proposed that scientific theories must include the possibility that they could be proved false by an experiment. Unfortunately, from a deductive point of view, no theory can be proved true or false. If an experiment fails to produce a predicted outcome, we know something went wrong. One possibility is, indeed, that the theory is false. But it could also be a problem in the experiment itself. Or the calculations that made the prediction. But there are other fundamental problems as well. When we make any calculations, we assume not only the theory to be tested. We also have to use many other auxiliary hypothesis about the world \cite{duhem89a,quine69}. When Newtonian Mechanics failed to predict the correct orbit of Uranus, the theory was not the one to blame. The actual error was the incomplete model of the Solar System. The initial calculations did not include the existence of Neptune. Current attempts to reconcile the movement of galaxies with their observed mass by adding dark matter terms do the same thing. Dark matter terms are proposed as an auxiliary hypothesis. Without them, the predictions for the behavior of galaxies would just be wrong. Maybe dark matter exists, maybe the theory is wrong. Experiments always leave something underdetermined. As a matter of fact, one could go on adding new hypothesis as needed to rescue basically any model. The new theory might become a convoluted model, but we might find ways to allow any theory to be rescued and still fit observation. The Ptolemaic description of the Solar System did something like that.

	Popper acknowledged that there is no deductive solid way to say a theory is correct. But there is also no way to determine a theory is wrong. We are left with induction for this task and the uncertainties that come with it. Luckily, while inductive methods can not tell us whether an idea is tue or not, they may still provide estimates of how plausible an idea is \cite{cox61a}. And it turns out that we can obtain clear rules for making inductions.  Based on a few simple and reasonable assumptions, it can be shown that assigning plausibilities can be done by following the rules of probability and changing our estimates when presented with new data by using Bayes formula \cite{jaynes03}. Induction, from a logical point of view, is the same as probability and Bayesian statistics.  
	
	This realization leads us to interesting consequences. And some of those consequences are well known but poorly understood. For example, we might be capable of making incredibly accurate guesses, but we can not really know. Some descriptions of the world will become extremely improbable, but that is not the same as false. Ideas can not be really rejected by induction \cite{martins16b}. Despite the deceptive name of some statistical tools, we only at best say some ideas are improbable. Indeed, the current replication crisis in areas as medicine and psychology have an important component of publication bias \cite{ioannidis05a}, but it is also caused by the widespread use of $p$-values \cite{gigerenzermarewski15a,trafimowmarks15a}. Statistics should indeed change its misleading terminology and stop talking about rejection \cite{martins16b}. What we can do, instead, is to rank ideas from more probable ideas to extremely improbable ones.

	A second consequence of using Bayesian methods is not so clear at first. Bayesian methods allow one to compare how the probability ratio of two theories plus their full set of auxiliary hypothesis changes based on data. While there are questions about prior choice, the method is straightforward. But that only applies to the very specific question of which set of theory plus hypothesis is more likely. It says nothing about the probability ratio of the theories themselves. And the final answer, while in probability form, is not really a probability. That is because all other possibilities, necessary to the full assessment, were excluded. In principle, to really obtain probabilities, Bayesian methods require that we use every possible theory and their variations \cite{fitelsonthomason08}. It actually requires a kind of theoretical omniscience \cite{martins05c} where all theories are known beforehand. The simplicity of Bayes formula is misleading. Using it correctly involves far more than the easy formula suggests.
	
	And this misleading simplicity makes reasoning based on Bayesian methods harder than it should be. Instead of following this path, I will base my analysis here on an inductive method that is equivalent to Bayesian methods, that is, Solomonoff induction \cite{solomonoff64a}. In it, theories are replaced by algorithms and probabilities become sums over successful algorithms. But the methods produce the same results. The only difference is that, while choosing Bayesian initial probabilities, the priors,  is not a simple task and still a matter of a lot of research, Solomonoff induction has no open problems. It comes with what can be seen as priors already chosen by the method itself.

	\subsection{Algorithms and Theories}
	
	While a Bayesian analysis of theory acceptance can teach us a lot, a few important details are not easy to notice in the Bayesian framework. It can be useful to use a different but equivalent induction framework, Solomonoff induction \cite{solomonoff64a}. Solomonoff induction basically provides the recipe for an algorithmic general method for induction that is equivalent to Bayes rule with a built-in choice of prior distributions \cite{hutter07a,rathmannerhutter11a}. It is also impossible to implement as it requires infinite computational resources. As a matter of fact, the same is true for Bayesian methods. The infinities are just hidden in the Bayesian case by the fact we can estimate odds ratios of two cases. And we can also calculate probabilities if we assume a very limited number of scenarios are all possible theories. But that assumption, while useful, is wrong. The ease that one can obtain incomplete and wrong analysis using Bayesian techniques is behind some of the criticism Bayesian methods have encountered. 
	
	Issues such as the problem of the old evidence \cite{garber83a} would not be real if we were not limited beings. The old evidence problem is a doubt about how to use old data when a scientist creates a new theory. And the only way to solve it is to remake all calculations you had previously done without the new theory now including it. This is usually not possible to do, but it is the correct way. After all, the order one creates theories should not matter, we must get the same result we would get if all theories were indeed known from the start .  Solomonoff induction requires in a much clearer way that all theories and ideas should be known beforehand. 
	
	What we get is that complete solutions to the problem of induction are incomputable \cite{solomonoff1996a}. We can obtain estimates from partially applying each method. But there is no way to estimate the error we might be making by disregarding parts of the complete calculation. There is no known way to assign an error to either the probabilities we get from partial Bayesian methods nor the predictions we get from partial Solomonoff induction.

	So, how does Solomonoff induction works? In simplified terms, assume we have a string $X$ that represents what we know about the world (or a given problem). For each program size $m$, we must generate all programs that have this size.  Some of these programs will have as output the string of data $X$, followed by what we call the predictions of that program. If the output of a given program does not give us $X$, ignore that program. If it does give us $X$, keep its prediction about the ``future'' observations. Each prediction we obtain from this Solomonoff machine is the weighted average of the predictions of every program that generates $X$. The weights are a function of program size $m$, given by $2^{-m}$. Repeat the procedure for every possible size $m$ and you have solved the problem of induction.
	
	This recipe is both perfect and impossible to follow. Given all the practical impossibilities involved in applying his set of rules, Solomonoff inductions is basically a gold standard. It tells us how we should do induction and it serves to enlighten us about the path to improve our estimates as well as point out what we might be doing wrong.   Given our natural reasoning shortcomings, the existence of a standard, despite impossible, can be a useful tool. It can help us understand better how we should evaluate and compare theories.

	In this case, of course, some effort to translate concepts is required. Solomonoff algorithms do not correspond to a theory. And they are not even equivalent to a theory and all the auxiliary hypothesis required to perform calculations. When we make a computational implementation of any theory, we sometimes have a theory that has probabilistic components. In this case, by simply changing the seed of the random number generator or the generator we can get different strings. But different random generator seeds still correspond to the same theory and auxiliary hypothesis. That is true even for completely deterministic theories.  That happens because all measurement is subject to errors. If we try to simulate the experiment that generated string $X$, we must estimate the uncertainty of the measurement process. And, just as with random number generation, other calculations that we implement tend to have more than one way to be implemented. The case of a single theory  and all needed auxiliary hypothesis we use to get a prediction still corresponds to a plethora of different algorithms.

	This huge set is actually needed, since, most of the time, the random generator will not provide draws that match the exact observed $X$. That will happen even if the theory was actually right. Here, precision will also play an important role. If the theory provides a very narrow distribution around the values actually in $X$, it is possible that several different realizations will return $X$. Wide distributions or distributions that are not centered around $X$, on the other hand, will more rarely turn our observed string, if ever. As a result, the theories that are indeed better will have more programs associated with them.  Their predictions will count more in the final average. Certainty is not achieved. But our predictions might be based with far more weight associated to one theory than its competitors.

	It is worth noticing that, while a theory will correspond to many different programs, there is no reason to expect that every program could be translated into a ``reasonable'' theory.  And some algorithms might, in principle, correspond to more than one theory.  If we just implement the final equations of two theories that predict the exact same behavior, a single algorithm may correspond to two or more theories. However, that does not mean that underdetermined theories will have the same set of algorithms. Implementation of the final dynamical equations might be identical, but there will be implementations where the consequences are calculated by the algorithm from the basic axioms. And those will be different.
	
	In the end, this whole process is equivalent to making a prediction using Bayesian methods. Larger weights translate to larger posterior probabilities. And the predictions come from a weighted average of the theories.

	\subsection{Pursue every theory}
	
	One thing should have become clear now. By averaging over every algorithm that produces the observed $X$, we are actually including each and every theory that is compatible with the observations. Even theories that are a little off will generate $X$ from algorithms with a specific choice of seed for a potent random generator. More serious disagreement can cause this to be so rare that the actual effect on the final average is close to null. But we still keep and inspect every possibility.

	Some theories will survive in many different algorithms.  Other theories will become quite rare. In that sense, Solomonoff induction is compatible with an assessment that theory $A$ is a much better description than theory $B$, much more probable. Those are equivalent ways to say the same thing. But both $A$ and $B$ survive. That means that our standard for correct induction  uses the whole space of possible theories. Therefore we must study the set of all possible theories as completely as we can. Once we do that, we will have theories that are far more probable given our data. And some that are so improbable that, aside a small technical abuse of language, we can call them false. It is likely we would end this process with more than one surviving theory that has a reasonable probability associated to it. But, since we are limited, we must start our analysis with a small set of theories. To correct that, Solomonoff induction suggest that we should continuously look for other theories that are compatible with our data. And keep each and everyone of them that are already compatible, for as long as they remain so.

	\section{What about String Theory?}
	
	An important consequence of a complete induction standard should be clear by now. If a new theory provides the same predictions as an older one, none of them can be discarded, if those predictions are compatible with the experiments. Instead, underdetermination means we will have to live with both versions. Excluding one has no solid justification. Despite our best hopes, we might never have certainty about the one theory of the universe. And it is perfectly reasonable to be forced to live with alternatives, if all alternatives describe our observations well. To desire one theory is a characteristic of our fallible minds. We want one true to believe and defend, so that we can stop thinking. But solid reasoning offers no such easy answers. The quest for unified theories can, of course, proceed. But we must understand that the final answer might be one unified theory as well as it might turn out to be many theories.
	
	Once we accept that, we can proceed. And there are still questions to answer. String Theory survives as a possibility we can not disregard, not without experiments that are, for now, impossible. Penrose metaphor of a scientist walking around an empty city and trying to find the one real beautiful place by looking for signs of aesthetically pleasing neighborhoods \cite{penrose05a} still provides interesting but probably unplanned illustrations of the problem. In the original version, there is only one building that we want to reach. In reality, there might be many places in the city that are worth visiting. Even if one is better than all the others, the only way to determine that is visiting them all and comparing.

	One question that remains is how much attention String Theory deserves. Or, in Bayesian terms, how probable it is that String Theory is the best alternative we have? An important argument that defends that it should be considered at least probable is based on the theoretical characteristics of the theory \cite{dawid06a,dawid13a}. But Bayesian methods only allow for probability change when we observe new data. Theoretical characteristics are not new data. And yet, the idea that a theory might be made more or less probable not only by data but also because of those theoretical characteristics has defenders.  It would be easy to identify such a defense with our desire to win arguments rather than seek for the truth. But it is worth asking if there might not be something hidden in those arguments.

	Taken at face value, the concept that anything other than new data can alter our evaluation is normatively wrong. While true, this statement is about the idealized version of the reasoning process. Limited humans might neglected to evaluate some relevant aspect of the problem. If later, we discover we have missed something, it might make sense to correct our estimates to account for the neglected facts.

	Indeed, some of the theoretical considerations proposed by Dawid \cite{dawid06a} can be understood as such a correction, while at least one of the arguments is wrong. The theoretical consideration that there is no other choice is false. First of all, there might be unconceived alternatives \cite{stanford01a} we have yet to find. And we already have other proposals for doing quantum gravity. Loop quantum gravity  is one example\cite{rovellismolin90}. More than that, if one looks only at the string of all data $X$ we have today, quantum field theory and general relativity seem to describe it all well, despite their incompatibilities. An algorithm where one of the two theories is used depending on the problem with a sensible rule to choose is, indeed, compatible with current $X$. While we might feel that is inelegant and wrong, we should know better than to use any feelings we might have. Normative rules exist exactly because our feelings often lead us astray. Of course, if there was only one algorithm that generated $X$ and we had access to every possible algorithm, we could say we had no choice. Realistically, if we knew of only one algorithm that generated $X$, we could still say that is all we have for now. But we are not there.

	The second theoretical argument used to defend String Theory is internal coherence. This argument goes like this. While String Theory was planned as a way to circumvent problems with infinities in quantum field theories, it has achieved much more. Gravity can be obtained as a consequence. String Theory can also help us understand better the concept of supersymmetry as well as black holes entropies. Indeed, as string theory seems capable of explaining more  and more about the structure and laws of the Universe, it does start to sound more like a solid theory. But our normative rules don't include the possibility of changing probabilities in this case.

	It is here that Solomonoff induction can provide a good illustration about what is going on. Matching the string $X$ with a given algorithm is a simple yes or no verification. The output either matches and the predictions of the algorithm are recorded with the proper weight, or the output does not match and the algorithm is discarded. The process of obtaining new data means adding to the string $X$. When we do that, more algorithms stop matching and are discarded, increasing the importance of those that remain. No considerations about the characteristics of the algorithms  are relevant at this point. Not when more data is obtained.

	But that is not true about the whole process. Indeed, one central part of the process is to assign weights to different algorithms. The longer they are, the less weight they have. One short algorithm may have a much larger weight or, in Bayesian terms, a larger prior probability. When we are debating theories about particles, forces, and gravity, our string $X$ includes all we know about each of those issues. If we need different theories to address each case and rules about when to use which one, that makes for a larger algorithm. On the other hand, if we can get $X$ from a small set of principles, this new set of principles can correspond to a smaller algorithm. A smaller algorithm means a larger weight in predictions. And that corresponds to a larger prior probability.

	From a normative point of view, this is no change in probability. In that sense, the criticism that we must not use theoretical considerations to update probabilities is technically correct. But we should have already included the theoretical considerations beforehand. Theoretical considerations should have influenced the prior distributions. Only then, we should move on and start learning from an expanding $X$, actually doing what we call updating probabilities. Unfortunately, as finite beings, this is not possible. We are always creating new theories. When that happens, we must repeat all the evaluation as if we had always known the theory from the beginning. This might not be feasible. Bayesian epistemologists call this issue the problem of the old evidence. It is quite clear what we should do in principle. It is not so clear how to actually solve the problem without the daunting task of processing every previous observation again. One practical suggestion might be that, as we remake our estimates, theoretical considerations can and should be taken into account. And, instead of entering in the priors, as theoretical considerations should, we may have to account for them later. As such, they will have an impact in the posterior estimates. That does not happen because theoretical characteristics can be used to update probabilities. It happens because we were unable to include those characteristics when we should have done so.

	The third argument used to support string theory is that its development reflects the successful development of the standard model. Both theories started from an attempt to solve technical problems with the previous descriptions. And it is clear that the standard model was quite successful at it, as its predictions have been confirmed by experiments. But there was a time in the past when we only had the predictions and no confirmation.  The community was still working on the technology to build large enough accelerators to check the model claims. The standard model was considered a valid approach then, and only later it was confirmed by experiments. Therefore, looking for a way to solve technical difficulties was considered then a valid procedure. The history of the standard model seem to suggest that, if you just find one solution to a technical problem, that solution might be correct.

	It may sound fair to ask for a similar treatment as the one that the standard model received. But that is no solid argument about the reliability of the theory in itself. It could simply mean the standard model was accepted too easily. In particular, normative solutions to the problem of induction include no comparison to human past behavior, even the behavior of scientists, and with good reason. So, it seems this argument should be properly ignored. It really does not fit our scheme of how a perfect induction must be performed. But, in reality, performing a perfect induction is an impossible task.

	We know we are unable to do the complete, perfect case. What this third argument does is to pose an interesting question. We do not have the full set of theories. And our resources to create new theories are limited. It can be a reasonable strategy to look for clues on which theoretical paths are more worth pursuing. While a heuristics about theoretical paths have no place in the idealized induction solution, it can still be very useful in the real life. Induction will tell us nothing here. Our normative solutions assume we know all theories. In this case, learning from successful cases in the past can be a good strategy. The third argument is basically a theory on which theories are more likely to be successful. The argument does not make string theory more probable. But it can make it a decent bet on where we should spend our efforts. As any heuristic, this one might fail. But if a heuristic is the best we have, it makes sense to use it and allow the string theory program to go on.

	\section{Conclusion}

	Human reasoning is flawed. We need mathematical and logical rules to be able to say with some degree of certainty that an idea might be true. Our tendency to defend the ideas we like instead of making impartial estimates means we need normative rules to replace our judgments in theory choice. Luckily, we have now solutions to the problem of induction. Bayesian statistics and the Solomonoff induction are those solutions. They are also equivalent. The only difference is the fact that Solomonoff induction comes with rules that, translated into the Bayesian method, correspond to a pre-choice of priors. What both accounts immediately show is that the question of whether a theory is scientific or not makes no sense. All theories are acceptable and should have their consequences calculated. From there we can rank them as more or less probable and that is it. Some theories will describe the data so badly that their probabilities will be ridiculously close to zero. But that still does not make those theories not scientific. It just makes them bad descriptions of the world. Not only string theory, but non-testable conjectures such as the existence of a multiverse or inflationary theories \cite{steinhardt14a} are valid scientific ideas. That is a very different statement from claiming any of those ideas has decent chances of being the best theory. For that, we do require better confirmatory evidence such as data. Some among us must learn to accept that untestable ideas such as the multiverse and inflation are valid theories. Others must accept that without experimental confirmation, they are not likely to be true. Unless they come as predictions from the majority of theories that are compatible with the data. Sadly, the space of all theories is an infinite space we are not capable of exploring fully. We are left with doubt. When underdeterminacy happens, competing theories survive and we can not discard any of them. Once we have some data, if there is still underdetermination, the corresponding theories survive with the same initial relative weight. To get rid of this prior, we can only keep on working until we find differences between the theories and ways to measure them. If that takes a long time and we feel frustrated, that says nothing about the theories. It is only about our desire to know the truth. That desire can be actually quite harmful to our ability to make sound judgments \cite{martins16a}.

	A question that needed an answer was if only data can provide confirmation. String theory seemed to get its support from theoretical arguments, instead of data. I have shown Solomonoff induction can help us understand better these questions.  Dawid's arguments \cite{dawid06a} may sound just plainly wrong from a purely Bayesian point of view. But by understanding how they stand in terms of Solomonoff induction, we were able to see their real weaknesses and strengths. Questions about the simplicity and power of a theory actually have an effect on the weight these theories have. In Bayesian terms, the characteristics of a theory can influence its prior probability. More powerful theories that can be described as smaller algorithms from where everything can be computed correspond to larger Solomonoff weights. They have larger priors. It makes sense to consider String Theory somehow validated by its theoretical characteristics. This should have happened before we start collecting any data. As theories appear after data is collected, we can only correct our analysis later.

	Finally, the idea that a strategy that provided us verifiable and good theories in the past might do the same again poses an interesting question. Our models of normative induction have nothing to say about this problem, as they assume all theories are known before any data is collected. As we are limited, it makes sense to use this kind of heuristic to evaluate which theoretical paths are more likely to produce good theories. While this heuristic might not make string theory more probable, it does suggest it is worth understanding it better.

	\section*{Acknowledgments}
	The author would like to thank the Funda\c{c}\~ao de Amparo \`a Pesquisa do Estado de S\~ao Paulo (FAPESP) for partial support to this research under grant 2014/00551-0.
	
	\bibliography{biblio}    

\begin{thebibliography}{10}

\bibitem{wigner60a}
Eugene Wigner.
\newblock The unreasonable effectiveness of mathematics in the natural
  sciences.
\newblock {\em Communications in Pure and Applied Mathematics}, 13(1), 1960.

\bibitem{ellissilk14a}
George Ellis and Joe Silk.
\newblock Scientific method: Defend the integrity of physics.
\newblock 516:321--323, 2014.

\bibitem{jaynes03}
E.T. Jaynes.
\newblock {\em Probability Theory: The Logic of Science}.
\newblock Cambridge, Cambridge University Press, 2003.

\bibitem{dawid13a}
Richard Dawid.
\newblock {\em String Theory and the Scientific Method}.
\newblock Cambridge University Press, 2013.

\bibitem{castelvecchi15a}
Davide Castelvecchi.
\newblock Feuding physicists turn to philosophy for help.
\newblock 528:446--447, 2015.

\bibitem{dawid06a}
Richard Dawid.
\newblock Underdetermination and theory succession from the perspective of
  string theory.
\newblock 73(3):298--322, 2006.

\bibitem{martins16b}
Andr\'e C.~R. Martins.
\newblock Stop the tests: Opinion bias and statistical tests.
\newblock Arxiv preprint: 1611.06545, 2016.

\bibitem{solomonoff64a}
R.~J. Solomonoff.
\newblock A formal theory of inductive inference. part i.
\newblock {\em Information and Control}, 7(1):1--22, 1964.

\bibitem{merciersperber11a}
Hugo Mercier and Dan Sperber.
\newblock Why do humans reason? arguments for an argumentative theory.
\newblock {\em Behavioral and Brain Sciences}, 34:57--111, 2011.

\bibitem{kahan13a}
Dan~M. Kahan.
\newblock Ideology, motivated reasoning, and cognitive reflection.
\newblock {\em Judgment and Decision Making}, 8:407--424, 2013.

\bibitem{oskamp65a}
Stuart Oskamp.
\newblock Overconfidence in case-study judgments.
\newblock {\em Journal of Consulting Psychology}, 29(3):261--265, 1965.

\bibitem{tsaietal08a}
Claire~I. Tsai, Joshua Klayman, and Reid Hastie.
\newblock Effects of amount of information on judgment accuracy and confidence.
\newblock {\em Organizational Behavior and Human Decision Processes},
  107:97–105, 2008.

\bibitem{christensenszalanskbusyhead81a}
Jay~J. Christensen-Szalanski and James~B. Bushyhead.
\newblock Physicians' use of probabilistic information in a real clinical
  setting.
\newblock {\em Journal of Experimental Psychology: Human Perception and
  Performance}, 7(4):928--935, 1981.

\bibitem{murphywinkler84a}
Allan~H. Murphy and Robert~L. Winkler.
\newblock Probability forecasting in meteorology.
\newblock {\em Journal of the American Statistical Association},
  79(387):489--500, 1984.

\bibitem{dawesetal89a}
Robyn~M. Dawes, David Faust, and Paul~E. Meehl.
\newblock Clinical versus actuarial judgment.
\newblock {\em Science}, 243:1668--1674, 1989.

\bibitem{proninetal06a}
Emily Pronin, Daniel~M. Wegner, Kimberly McCarthy, and Sylvia Rodriguez.
\newblock Everyday magical powers: The role of apparent mental causation in the
  overestimation of personal influence.
\newblock {\em Journal of Personality and Social Psychology}, 91(2):218--231,
  2006.

\bibitem{martins16a}
Andr\'e C.~R. Martins.
\newblock Thou shalt not take sides: Cognition, logic and the need for changing
  how we believe.
\newblock {\em Frontiers in Physics}, 4(7), 2016.

\bibitem{popper59}
K.~Popper.
\newblock {\em The Logic of Scientific Discovery}.
\newblock London, Hutchinson, 1959.

\bibitem{duhem89a}
Pierre Duhem.
\newblock {\em La th\'eorie physique: Son Objet et sa Structure}.
\newblock 1989.

\bibitem{quine69}
W.V. Quine.
\newblock Epistemology naturalized.
\newblock In {\em Ontological Relativity and Other Essays}. New York, Columbia
  University Press, 1969.

\bibitem{cox61a}
R.~T. Cox.
\newblock {\em The Algebra of Probable Inference}.
\newblock John Hopkins University Press, 1961.

\bibitem{ioannidis05a}
J.~P.~A. Ioannidis.
\newblock Contradicted and initially stronger effects in highly cited clinical
  research.
\newblock {\em Journal of the American Medical Association}, 294:218--228,
  2005.

\bibitem{gigerenzermarewski15a}
Gerd Gigerenzer and Julian~N. Marewski.
\newblock Surrogate science: The idol of a universal method for scientific
  inference.
\newblock {\em Journal of Management}, 41(2):421--440, 2015.

\bibitem{trafimowmarks15a}
David Trafimow and Michael Marks.
\newblock Editorial.
\newblock {\em Basic and Applied Social Psychology}, 37:1--2, 2015.

\bibitem{fitelsonthomason08}
B.~Fitelson and N.~Thomason.
\newblock Bayesians sometimes cannot ignore even very implausible theories
  (even ones that have not yet been thought of).
\newblock {\em Australasian Journal of Logic}, 6:25--36, 2008.

\bibitem{martins05c}
Andr\'e C.~R. Martins.
\newblock Theoretical omniscience: Old evidence or new theory.
\newblock PhilSci Preprint at
  http://philsci-archive.pitt.edu/archive/00002458/, 2005.

\bibitem{hutter07a}
Marcus Hutter.
\newblock On universal prediction and bayesian confirmation.
\newblock 384:33--48, 2007.

\bibitem{rathmannerhutter11a}
Samuel Rathmanner and Marcus Hutter.
\newblock A philosophical treatise of universal induction.
\newblock {\em Entropy}, 13(6):1076, 2011.

\bibitem{garber83a}
D.~Garber.
\newblock Old evidence and logical omniscience in bayesian confirmation theory.
\newblock In J.~Earman, editor, {\em Testing Scientific Theories}, volume~X of
  {\em Minnesota Studies in the Philosophy of Science}. University of
  Minneapolis Press, 1983.

\bibitem{solomonoff1996a}
Ray Solomonoff.
\newblock Does algorithmic probability solve the problem of induction.
\newblock {\em Information, Statistics and Induction in Science}, pages 7--8,
  1996.

\bibitem{penrose05a}
Roger Penrose.
\newblock {\em The Road to Reality}.
\newblock Vintage Books, 2005.

\bibitem{rovellismolin90}
Carlo Rovelli and Lee Smolin.
\newblock Loop space representation of quantum general relativity.
\newblock {\em Nuclear Physics B}, 331(1):80 -- 152, 1990.

\bibitem{stanford01a}
P.~Kyle Stanford.
\newblock Refusing the devil's bargain: What kind of underdetermination should
  we take seriously?
\newblock 68:S1--S12, 2001.

\bibitem{steinhardt14a}
Paul Steinhardt.
\newblock Big bang blunder bursts the multiverse bubble.
\newblock 510:9, 2014.

\end{thebibliography}
	\bibliographystyle{unsrt}

\end{document}